\documentclass[aip,reprint,floatfix]{revtex4-1}
\usepackage{amsfonts,amssymb,amsmath,array}

\usepackage{dcolumn}
\usepackage{bm}
\usepackage[final]{graphicx}
\usepackage{graphicx}
\usepackage{epstopdf}
\graphicspath{{./Figures/}} %Setting the graphicspath

\usepackage[usenames]{color} %Per permettere la colorazione dei caratteri 

\usepackage{soul} %per cancellare le frasi

\begin{document}

\title{Correlated escape of active particles across a potential barrier}

\author{Lorenzo Caprini}
\affiliation{Scuola di Scienze e Tecnologie, Universit\`a di Camerino,  
Via Madonna delle Carceri, I-62032, Camerino, Italy} 
\affiliation{Heinrich-Heine-University of D\"usseldorf, Universit\"atsstrasse 1, D\"usseldorf, Germany}
  
\author{Fabio Cecconi}
\affiliation{Istituto dei Sistemi Complessi (CNR), Via Taurini 19, I-00185 Roma, Italy}

\author{Umberto Marini Bettolo Marconi}
\affiliation{Scuola di Scienze e Tecnologie, Universit\`a di Camerino,  
Via Madonna delle Carceri, I-62032, Camerino, Italy}

\begin{abstract}
We study the dynamics of one-dimensional active particles confined in a double-well potential, focusing on the escape properties of the system, such as the mean escape time from a well. 
We first consider a single-particle both in near and far-from-equilibrium regimes by varying the persistence time of the active force and the swim velocity.  
A non-monotonic behavior of the mean escape time is observed with the persistence time of the activity, revealing the existence of an optimal choice of the parameters favoring the escape process.
For small persistence times, a Kramers-like formula with an effective potential obtained within the Unified Colored Noise Approximation is shown to hold.
Instead, for large persistence times, we developed a simple theoretical argument based on the first passage theory which explains the linear dependence on the escape time with the persistence of the active force.
In the second part of the work, we consider the escape of two active particles mutually repelling. 
Interestingly, the subtle interplay of active and repulsive
forces may lead to a correlation between particles favoring the simultaneous jump across the barrier. 
This mechanism cannot be observed in the escape process of two passive particles. 
Finally, we find that, in the small-persistence regime, the repulsion favors the escape, like in passive systems, in agreement with our theoretical predictions, while for large persistence times, the repulsive and active forces produce an effective attraction which hinders the barrier crossing.
\end{abstract}

\maketitle

%\tableofcontents

%%%%%%%%%%%%%%%%%%%%%%%%%%%%%%%%%%%%%%%%%%%%%%%%%%%%%%%%%%%%%%%%%%%
\section{\label{sec:intro} Introduction}
%%%%%%%%%%%%%%%%%%%%%%%%%%%%%%%%%%%%%%%%%%%%%%%%%%%%%%%%%%%%%%%%%%%

Nowadays, biological systems, such as bacteria and cells, or some specific classes of colloids are classified as {\it{active}}~\cite{bechinger2016active, marchetti2013hydrodynamics, elgeti2015physics}. They distinguish from {\it{passive}} systems for a plethora of interesting phenomena experimentally observed which opens the way to many intriguing medical and engineering applications~\cite{gompper20202020}.
Active systems often accumulate near obstacles~\cite{mino2018coli, maggi2015micromotors, angelani2009self} and boundaries~\cite{vladescu2014filling, mok2019geometric, caprini2019active} and display collective phenomena such as living-clusters \cite{palacci2013living, mognetti2013living}, motility induced phase separation~\cite{solon2015pressure, buttinoni2013dynamical, petrelli2018active, bialke2015active, van2019interrupted, caporusso2020motility} and spatial velocity correlations~\cite{peruani2012collective, garcia2015physics, caprini2020hidden, henkes2020dense}. These properties have been reproduced with the help of coarse-grained stochastic models that neglect the biological or chemical origin of the activity 
in favor of an additional degree of freedom, simply referred to as {\it{active force}}~\cite{fodor2018statistical, shaebani2020computational}.
This ingredient guarantees the time-persistence of single-trajectory experimentally observed that is recognized to be a fundamental hallmark
of active systems.

Recently, the behavior of active systems confined in thin geometries or by external potentials has been a matter of intense investigation~\cite{szamel2014self, das2018confined, caprini2019activity, malakar2020steady, hennes2014self, rana2019tuning, holubec2020active} through both experimental and numerical studies.
For instance, active colloids could be confined in external potential by magnetic or optical tweezers~\cite{militaru2021escape} and recently by using acoustic traps~\cite{takatori2016acoustic}, while the confinement for Hexbug particles, i.e. macroscopic self-propelled toy robot, could be simply achieved through a parabolic dish~\cite{dauchot2019dynamics}.  
Some approximate theoretical treatments have been formulated to predict the statistical properties of active systems~\cite{wittmann2017effective}.
These include: the diffusion properties in complex environments \cite{caprini2020diffusion, breoni2020active}, the probability distribution function (displaying strong deviation from Boltzmann profiles)~\cite{fodor2016far, marconi2017heat}, the pair correlation functions~\cite{marconi2016effective, wittmann2016active}, the pressure and the  surface tension~\cite{wittmann2019pressure}. % and the shape of the spatial velocity correlations~\cite{caprini2021spatial}.
However, these methods usually work in specific regimes of parameters and, in some cases, break down in regimes of strong (e.g. persistent) activity.
This failure is found in the case of the active version of a celebrated problem of equilibrium (e.g. passive) statistical mechanics: the escape from a potential barrier, also known as Kramers problem~\cite{hanggi1990reaction, mel1991kramers}.
In this context, a paradigmatic case that has received much attention in the literature of passive particles concerns the escape in a double-well potential.
In the active case, some analytical results have been obtained in near equilibrium regimes~\cite{sharma2017escape, geiseler2016kramers, Scacchi2018mean}, where the average escape time can be analytically predicted by taking advantage of equilibrium-like approximations.
Subsequently, most of the studies have been focused on far-from-equilibrium regimes (large swim velocities and/or large persistence times) showing behaviors without passive counterparts: for instance, Woillez et. al. found that the escape time of active particles is affected by the whole shape of the potential and not only by the height of the potential barrier~\cite{woillez2019activated}.
In addition, some peculiar properties of the escape mechanism in the large persistence regime have been discussed in Ref.~\cite{caprini2019activeescape}: a bifurcation-like scenario in the position-velocity phase space emerges. A suitable theory for the escape rate in this regime has been derived by using large deviation techniques holding in the limit of infinite persistence time~\cite{woillez2020nonlocal}, while, in the same regime of parameters, Fily obtained an approximate expression for the probability distribution~\cite{fily2019self}.
Finally, Debnath et. al.~\cite{Debnath2018Activated} focused on the escape time
in two dimensions and in the presence of hydrodynamic interactions of an active Brownian particle carrying a passive cargo.

The interest in the active escape processes goes beyond the paradigmatic case of the double-well potential as testified by the studies on the escape from other potential shapes~\cite{scacchi2019escape, dhar2019run}, for instance, harmonic potentials~\cite{wexler2020dynamics, gu2020stochastic} with interesting applications for the active version of the trap model~\cite{woillez2020active} or the behavior of active particles in rugged energy landscapes~\cite{chaki2020escape}. 
It is important to remark that theoretical results for the mean escape time in the presence of general potentials have been only derived in the limit of small active force~\cite{Benjamin2021First}.
Finally, recent works have studied the escape of active particles from thin openings of confining geometries, such as disks~\cite{olsen2020escape, paoluzzi2020narrow, Biswas2020First}, for their potential applications in biological processes~\cite{schuss2007narrow}. 
The average escape time has been investigated also in simpler geometries such as one-dimensional channels~\cite{locatelli2015active}, open-wedge channels~\cite{caprini2019transport} and channels with bottlenecks~\cite{ghosh2014communication} suitable to model the escape of living organisms from biological pores.

In this paper, we focus on the escape properties of active particles evaluating both small, intermediate and large persistence and find the occurrence of an optimal persistence favoring the escape from the potential barrier.
In the second part of the work, the same problem is addressed by considering a system formed by two interacting active particles to assess the effect of repulsive interactions on the escape process.
The paper is structured as follows: In Sec.~\ref{sec:model}, we introduce the model used to simulate the dynamics of active particles, while in Secs.~\ref{sec:oneparticle} and~\ref{sec:twoparticle}, we study the escape problem for one and two interactive particles, respectively.
In the final section, we present the conclusions.

%%%%%%%%%%%%%%%%%%%%%%%%%%%%%%%%%%%%%%%%%%%%%%%%%%%%%%%%%%%%%%%%%%%
\section{\label{sec:model} Model}
%%%%%%%%%%%%%%%%%%%%%%%%%%%%%%%%%%%%%%%%%%%%%%%%%%%%%%%%%%%%%%%%%%%

%---------------------------- Fig.1 ----------------------------------
\begin{figure*}[t!]
\includegraphics[width=0.85\textwidth]{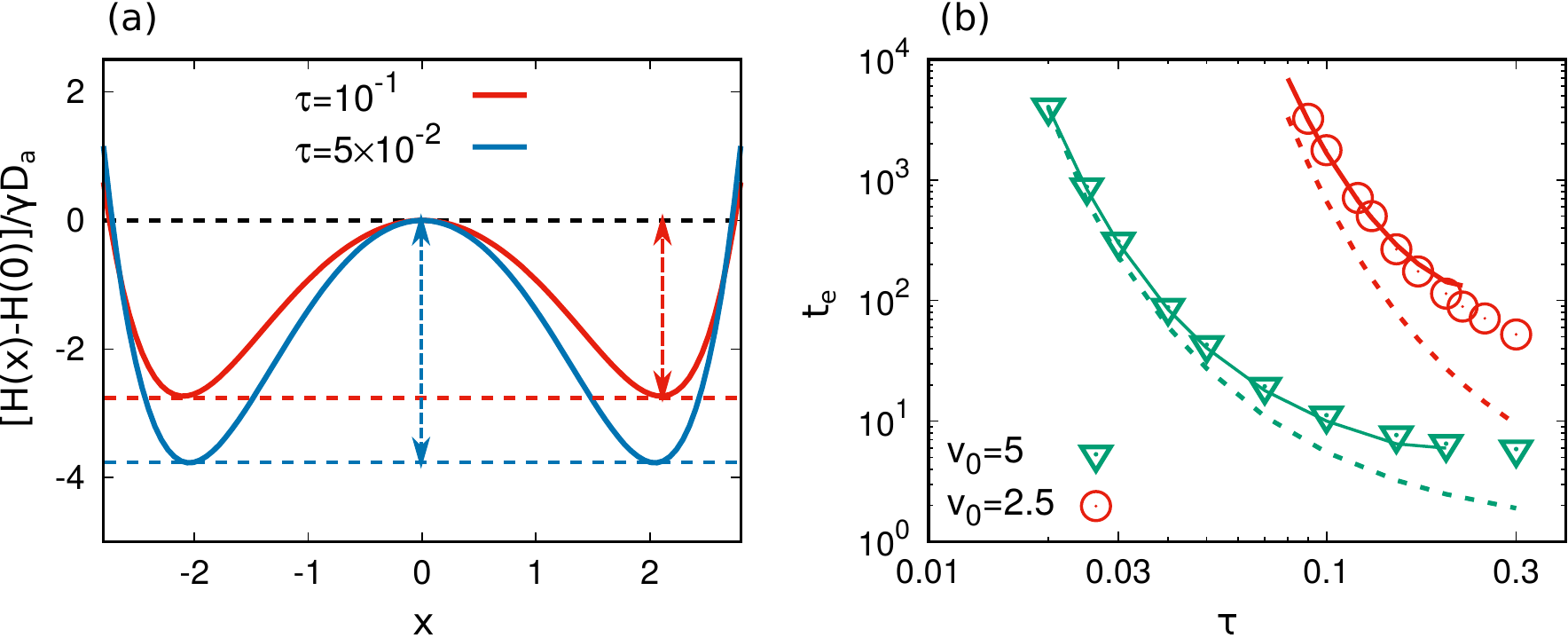}
\caption{Panel~(a): Effective potential, $(H(x)-H(0))/\gamma D_a$, where $D_a=v_0^2\tau$, for $\tau=10^{-1}$ (red curve) and $\tau=5\times10^{-2}$ (blue curve). Horizontal dashed lines are guides for the eyes, marking the value of the maximum and minima of the effective potential, while vertical dashed arrows measure the height of the effective potential barrier, $H_b/(D_a \gamma)$.
Panel~(b): average escape time, $t_e$, as a function of the persistence time $\tau$ for two different values of the swim velocity $v_0$. Colored points are the results of numerical simulations, solid lines are obtained from Eq.~\eqref{eq:kramersUCNA} and dashed ones from Eq.~\eqref{eq:kramersPassive}. 
We remark that the prediction~\eqref{eq:kramersUCNA} is shown for those values of $\tau$ for which the UCNA prediction is defined.
The other parameters of the simulations are: $W_0=1$ and $a=2$.
\label{fig:Time_smalltau}}
\end{figure*}
%---------------------------------------------------------------------

We consider the dynamics of active particles confined in an external double-well potential of the form: 
\begin{equation}
\label{eq:doublewell}
W(x) = \dfrac{W_0}{4}(x^2-a^2)^2 \,.
\end{equation}
%where the constants $V_0$ and $a$ determine the amplitude of the potential barrier and the distance between the potential minima, respectively.
The profile of $W(x)$ is characterized by two minima at $x=\pm a$ separated by a potential barrier at $x=0$ of 
height $W_b = W_0 a^4/4$. 
The dynamics is made {\it active} by including a stochastic force, $f^a_i$, in the evolution for the particle position $x_i$. 
$f^a_i$ is generated by an Ornstein-Uhlenbeck process and is characterized by a persistence time $\tau$ and a variance $v_0$. 
This choice corresponds to the so-called active Ornstein-Uhlenbeck particles (AOUP) model~\cite{berthier2017active, mandal2017entropy, wittmann2017effective, dabelow2019irreversibility, caprini2021inertial, martin2021statistical, nguyen2021active}, and reproduces the typical phenomenology of active particles~\cite{fodor2016far, caprini2020time, maggi2021universality}. 
The AOUP has been often employed as an approximation for other popular active models~\cite{fily2012athermal, farage2015effective, caprini2019comparative} or  to describe the behavior of a colloidal particle in a bath of active particles~\cite{wu2000particle, maggi2014generalized, chaki2019effects}. 
The equations of motion for $N$ overdamped AOUP with position $x_i$ are given by:   
\begin{subequations}
\label{eq:activedynamics}
%\begin{eqnarray}
\begin{align}
\gamma\dot{x}_i&=   f^a_i - W'(x_i) + F_i \\
%\end{equation}
%\begin{equation}
\label{eq:activedynamics_b}
\tau\dot{f}^a_i &=  -f^a_i + \gamma v_0 \sqrt{2\tau} \xi_i \,,
\end{align}
%\end{eqnarray}
\end{subequations}
where $\xi_i$ is a white noise with zero average and unit variance such that $\langle \xi_i(t) \xi_j(0) \rangle = \delta_{ij}\delta(t)$ and $\gamma$ is the friction coefficient.
%\begin{equation}
%U(x_1,x_2) = W(x_1) + W(x_2) + U(|x_2-x_1|)
%\end{equation}
Finally, the last force term in Eq.~\eqref{eq:activedynamics}, namely $F_i=-\partial_{x_i} U_{tot}$, is due to the repulsive interactions between neighboring particles and models volume exclusion.
This force corresponds to the potential $U_{tot}=\sum_{i<j} U(|x_j-x_i|)$, where $U$ is a Weeks-Chandler-Andersen (WCA) potential of the form: 
\begin{equation}
U(r) =
\begin{cases} 
4\epsilon\bigg[\bigg(\dfrac{\sigma}{r}\bigg)^{12} - 
2\bigg(\dfrac{\sigma}{r}\bigg)^6 +1\bigg] & r\le \sigma \\
0 & r> \sigma 
\end{cases} 
\end{equation}
where $r = |x_i - x_j|$ is the distance between the two particles, $\sigma$ is the nominal particle diameter, and $\epsilon$ the typical energy scale of the interaction;  
for simplicity, we set $\sigma=1$ and $\epsilon=1$.

%%%%%%%%%%%%%%%%%%%%%%%%%%%%%%%%%%%%%%%%%%%%%%%%%%%%%%%%%%%%%%%%%%%%%%
\section{One-particle active escape problem
\label{sec:oneparticle}}
%%%%%%%%%%%%%%%%%%%%%%%%%%%%%%%%%%%%%%%%%%%%%%%%%%%%%%%%%%%%%%%%%%%%%
Before delving into the case of two particles, we consider a single particle, i.e. the dynamics~\eqref{eq:activedynamics} with $N=1$. In this section, for simplicity, we omit Latin subscripts.

In the one-particle case, Refs.~\cite{caprini2019activeescape, woillez2020nonlocal} have been already shown that the jump mechanism from a potential well to the other is strongly affected by the persistence time $\tau$ which can change also qualitatively the dynamical picture of the escape process.
Roughly speaking, we can distinguish two limiting mechanisms depending on the values of $\tau$ considered: i) the regime of small $\tau$, such that $f^a$ relaxes faster than the particle position, $x$,
and ii) a regime of large persistence, such that $f^a$ relaxes slower than $x$.

%$x_m=a+ \tau***$
%$(H(a)-H(0))/(D_a\gamma) = \dfrac{W_0 a^4}{4 v_0^2\tau \gamma} + \dfrac{3 W_0 a^2 }{v_0^2\gamma v_0^2 \gamma}$

%=======================================================================
\subsection{Small persistence regime}
%=======================================================================
When $f^a$ relaxes faster than $x$ (namely, in the regime of small $\tau$), the system is near the equilibrium and the unified colored noise approximation (UCNA) applies~\cite{jung1987dynamical, maggi2015multidimensional, wittmann2017effective}. 
This means that the role of the active force can be recast onto an effective potential with an effective diffusion coefficient $D_a=v_0^2 \tau$.
In practice, the active particle behaves as a Brownian-like particle described by the probability distribution:
\begin{equation}
\label{eq:UCNAprob}
p(x) \propto \exp[-H(x)/D_a\gamma] \,,
\end{equation}
%whose statistical properties are contained in the effective Hamiltonian (potential)~\cite{maggi2015multidimensional}: 
where $H(x)$ is the following effective potential~\cite{maggi2015multidimensional}: 
\begin{equation}
\label{eq:HamiltonianUCNA}
H(x)=W(x)+\frac{\tau}{2\gamma} (W'(x))^2 - D_a \gamma\log{\left( 1+\frac{\tau}{\gamma} W''(x) \right)} \,,
\end{equation}
where the prime denotes the spatial derivative. 
$H(x)$ can be interpreted as the Hamiltonian of a passive particle depending on $W(x)$ and its derivatives. 
The extra terms in Eq.~\eqref{eq:HamiltonianUCNA} maintain the symmetric two-well structure 
(see Fig.~\ref{fig:Time_smalltau}~(a) illustrating $H(x)/D_a\gamma$ for two different values of $\tau$),
but shift the positions of the two minima, $|x_m|$, and change the height of the effective potential barrier, $H_{b}=H(0)-H(|x_m|)$.
To first order in $\tau$, we obtain:
\begin{subequations}
\label{eq:effective_potential_parameters}
\begin{align}
\label{eq:effectivepotentialminimum}
&|x_m| \simeq a \left(1+ 3 \frac{D_a \tau}{a^2}\right) \\  
\label{eq:effectivepotentialbarrier}
&H_b \simeq \frac{a^4 W_0}{4}\bigg(1 + 12 \frac{D_a\tau}{a^2}\bigg) \,.
\end{align}
\end{subequations}
The height of the potential barrier increases with the activity, thus one would expect that a larger activity hinders the passage from one well to the other. On the contrary, since the key factor controlling the escape dynamics is the ratio $H_b/D_a \gamma=H_b/(v_0^2\tau\gamma)$ which decreases with $\tau$ (and $v_0$), one finds that the active force favors the escape process (as shown in Fig.~\ref{fig:Time_smalltau}~(a)).
In summary, in the small persistence regime, the jump process resembles its passive counterpart and the average escape time, $t_e$, can be estimated by applying the Kramers formula~\cite{gardiner1985handbook} by considering the effective potential~\eqref{eq:HamiltonianUCNA}:
\begin{equation}
\label{eq:kramersUCNA}
t_e \approx \frac{2\pi}{\sqrt{H''(|x_m|)\;|H''(0)|}} \exp\left({\frac{H_b}{D_a\gamma}}\right) \,,
\end{equation}
where $H''$ denotes the second derivative of Eq.~\eqref{eq:HamiltonianUCNA} and, as usual, formula~\eqref{eq:kramersUCNA} holds when $H_b/(D_a\gamma)\gg 1$.
As a matter of fact, in the regime of small $\tau$, the leading term in the expression~\eqref{eq:HamiltonianUCNA} is simply the potential $W(x)$ while the remaining terms provide small corrections whose relevance increases as $\tau$ grows.
As a consequence, for $\tau$ small enough (when the $O(\tau)$ are negligible in the expression for $H(x)$), Eq.~\eqref{eq:kramersUCNA} reduces to: 
\begin{equation}
\label{eq:kramersPassive}
t_e \sim \exp{\left[W_b /(v_0^2 \tau \gamma)  \right]}\,,
\end{equation}
which mainly depends on the height $W_b$ of the potential barrier, showing that $t_e$ exponentially decreases when $\tau$ or $v_0$ are increased.

Predictions~\eqref{eq:kramersUCNA} and~\eqref{eq:kramersPassive} have been checked numerically, as shown in Fig.~\ref{fig:Time_smalltau}(b), where the escape time $t_e$ is plotted as a function of $\tau$ for two different values of $v_0$.
As expected from the naive passive-like formula~\eqref{eq:kramersPassive}, $t_e$ decreases as $\tau$ grows, and similarly it becomes smaller as $v_0$ increases.
However, a quantitative agreement in a wide $\tau$-range is achieved
only by using the UCNA result \eqref{eq:kramersUCNA} whereas Eq.~\eqref{eq:kramersPassive} fails at larger values of $\tau$.
%---------------------------- Fig.2 ---------------------------------
\begin{figure*}[t!]
\includegraphics[width=0.9\textwidth]{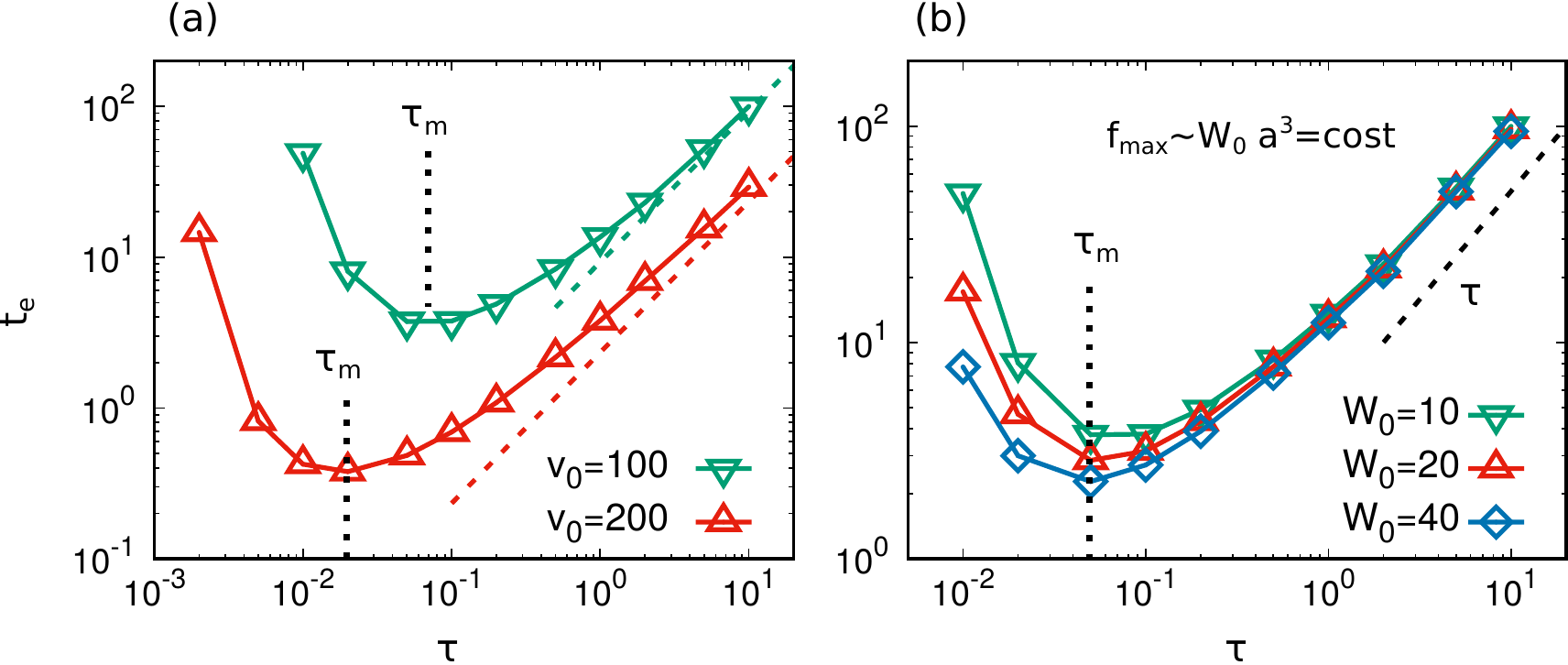}
\caption{Average escape time, $t_e$, as a function of the persistence time $\tau$ for two different values of the swim velocity $v_0=10^2, 2\times 10^2$ (with a fixed potential) in panel (a) and by keeping fixed $v_0$ and by changing the potential parameter in panel (b). In particular, in the latter case, $t_e$ is shown for three different values of the potential strength $W_0=10, 20, 40$ by keeping fixed the maximal force exerted in the flex point, $f_{M} \sim W_0 a^3$.
The colored dashed lines in panel (a) are obtained by using Eq.~\eqref{eq:prediction_te_largetau} with the values of $v_0$ indicated in the legend. Instead, the dashed black line in panel (b) qualitatively shows the linear scaling with $\tau$.
Finally, the dotted black lines are eye-guides which mark the value of $\tau_{m}$
The other parameters of panel (a) are: $W_0=10$, $a=\sqrt{10}$, while the other parameters of panel (b) are: $v_0=10^2$ and $a$ scaling accordingly to the value of $W_0$ such that $W_0 a^3=10^{5/2}$.
\label{fig:LargeTau}}
\end{figure*}
%---------------------------------------------------------------------

%==================================================================
\subsection{Large persistence regime}
%==================================================================
In the case where $f^a$ relaxes slower than $x$, that is in a regime of large persistence,
we can identify a jump mechanism differing from that of a passive particle~\cite{caprini2019activeescape}.
In fact, if $\tau$ exceeds a certain threshold, the UCNA distribution~\eqref{eq:UCNAprob} 
becomes ill-defined as the argument of the logarithm
$$
\log{\left( 1+\frac{\tau}{\gamma} W''(x) \right)}
$$
in Eq.~\eqref{eq:HamiltonianUCNA} becomes negative for certain values of $x$. 
At large enough value of $\tau$, this occurs because $W''(x)<0$ for $|x|<a/\sqrt{3}$.
Nevertheless, even for large $\tau$, as shown in Ref.~\cite{caprini2019activeescape}, around the potential minima the particle distribution can be fairly well represented by $p(x) \propto e^{-H(x)/D_a\gamma}$. 
%
%Instead, a different scenario is observed in the space region around the potential maximum.
%Specifically, the conditional distribution of the velocity, $v=\dot{x}$, at a fixed position, $p(v|x)$, has a Gaussian shape in the wells but  displays strong non-Gaussianity developing a bimodal character when $x$ is in the barrier region. 
Moreover, Ref.~\cite{caprini2019activeescape} shows that, in regime of large $\tau$, the jump process occurs almost deterministically when the modulus of the active force $|f_a|$ exceeds the threshold $f_{M} =2 W_0 a^3/(3 \sqrt{3})$ corresponding to the maximal force exerted by the double-well (i.e. the modulus of the force evaluated at the inflection points $\pm a/\sqrt{3}$).
In particular, jumps from the left (right) well to the right (left) one occur when $f_a>f_{M}$ ($f_a<-f_{M}$).
These results have been formalized using large-deviation techniques in Ref.~\cite{woillez2020nonlocal}, where an asymptotic expression for the distribution function, $p(x)$, has been derived in the limit $\tau \to\infty$.

To understand how the escape properties are modified by a persistent active force, we study the average escape time, $t_e$, as a function of $\tau$ and explore values for which UCNA does not hold. 
In particular, Fig.~\ref{fig:LargeTau}~(a) shows $t_e$ vs $\tau$ for different values of the swim velocity, $v_0$.
For each value of $\tau$, the larger $v_0$, the larger the value of $t_e$.   
Instead, as a function of $\tau$ a non-monotonic behavior is observed.
After a first decrease, which resembles the Kramers-like behavior~\eqref{eq:kramersPassive} described in the small $\tau$ regime, 
the escape time drops to a minimum for $\tau=\tau_m$, till to increase at larger values of $\tau$.
This means that, for a given potential set-up and swim velocity, one can identify an optimal value of the persistence time ($\tau_m$) favoring the jump process.
The above scenario can be explained through a simple argument based on the interplay between persistence length $\ell = v_0 \tau$ and the distance between maximum and minimum of the potential, $a$.
When $\ell \ll a$, the active force can change direction during the barrier climbing. In this regime, we have already seen in Fig.~\ref{fig:Time_smalltau} that $t_e$ decreases as $\tau$ grows.
Instead, when $\ell \gg a$, the escape occurs almost deterministically when the active force overcomes $f_{M}$ and $f^a$ has a little chance to reverse the direction during the barrier crossing, at variance with the small-$\tau$ regime where $f^a$ can invert its direction many times. Thus, for $\ell \gg a$, a barrier crossing is simply related to a first passage for the process~\eqref{eq:activedynamics_b}: the larger $\tau$, the larger the time waited for 
the occurrence of a value of the active force such that $f_a>f_{M}$, implying a larger $t_e$.
As a consequence, we expect the presence of a minimum in the intermediate regime, say $\ell=v_0 \tau \sim a$.
The argument explains the behavior of $\tau_m$ with the swim velocity: indeed, as shown in Fig.~\ref{fig:LargeTau}~(a), $\tau_m$ decreases with $v_0$ in agreement with the scaling:
$$
\tau_m \sim \frac{a}{v_0} \,. 
$$
We remark that the non-monotonic behavior with $\tau$ has been already observed in other observables of this system, such as the entropy production~\cite{dabelow2021irreversible} or the integrated linear response to a small perturbation, introduced to test the breakdown of the detailed balance~\cite{caprini2021fluctuation}.

As reported by Fig.~\ref{fig:LargeTau}~(a), the increase of $t_e$ with $\tau$ for $\tau>\tau_m$ is quite slow and displays an algebraic growth, which roughly approaches a linear dependence $t_e \sim \tau$ in the regime $\tau\gg\tau_m$.
This qualitative observation can be supported by a theoretical argument holding in the limit $\tau\to\infty$. 
As already discussed, since for $\tau\gg \tau_m$, a jump occurs only when $f^a>f_{M}$ (or $f^a<-f_{M}$), we can identify $t_e$ as the typical time taken by $f^a$ to reach $f_{M}$ (or $-f_{M})$, which is nothing but the first-passage problem of an Ornstein-Uhlenbeck process. 
This is briefly reviewed in Appendix~\ref{app:derivation} (see also Ref.~\cite{sato1977evaluation}) and provides the following prediction:
\begin{equation}
\begin{aligned}
\label{eq:prediction_te_largetau}
%t^{\tau\to\infty}_e&\approx \tau \sqrt{\pi} \int_0^{\phi} 
%du\;e^{u^2}\left( 1 + \text{Erf}\,[u]\right) \\
\lim_{\tau\to\infty} \, t_e \approx \tau \sqrt{\pi} \int_0^{\phi} 
du\;e^{u^2}\left( 1 + \text{Erf}\,[u]\right)  \,,
\end{aligned}
\end{equation}
where, $\phi=f_{M}/\sqrt{2}v_0\gamma$ is a dimensionless force and $\text{Erf}[\,\cdot\,]$ indicates the error function. % and, in the second line, we have applied the saddle point method which holds for $f_{M}\ll v_0\gamma$.
As shown in Fig.~\ref{fig:LargeTau}~(a) (see the dashed colored lines) the prediction~\eqref{eq:prediction_te_largetau} is in fair agreement with data (except for the presence of an extra factor $\sqrt{2}$ which is missed by our theory) and, in particular, allows us to explain the behavior $t_e\propto\tau$ observed by simulations for $\tau \gg \tau_m$.
In addition, the saddle-point method, applied to Eq.~\eqref{eq:prediction_te_largetau} for $f_{M}\ll v_0\gamma$, yields the expression
\begin{equation}
\label{eq:pred_largetau_saddlepoint}
\lim_{\tau\to\infty} t_e\approx  \tau\frac{f_M}{v_0\gamma} \sqrt{\frac{\pi}{2}} \exp\Bigl(\dfrac{f_{M}^2}{2v_0^2\gamma^2}\Bigr) \,,
\end{equation}
which coincides with the result of Ref.~\cite{woillez2020nonlocal} except for the prefactor depending on $\tau$, that has not been derived by Woillez et. al..
From Eqs.~\eqref{eq:prediction_te_largetau} and~\eqref{eq:pred_largetau_saddlepoint}, the difference with the passive escape problem is contained in the scaling of $t_e$ with the parameter of the potential.
Indeed, the average escape time of a passive Brownian particle mainly depends on the height of the potential barrier which scales as $W_0 a^4/4$ for the double-well given by Eq.~\eqref{eq:doublewell}.
Instead, in the active persistent case, in particular in the regime of large $\tau$ such that $\ell \gg a$ (so that $\tau \gg \tau_m$), the escape time depends on the value of the maximal force $f_{M} =2 W_0 a^3/(3 \sqrt{3})$ experienced along the barrier climbing.
This scaling is checked in Fig.~\ref{fig:LargeTau}~(b), where $t_e$ is shown as a function of $\tau$ at fixed $v_0$, by varying $a$ and $W_0$ such that $W_0 a^3=\mbox{const}$ to keep $f_{M}$ constant.
In particular, we plot three different values of $W_0$  showing the collapse of the curves when $\tau \gg \tau_m$.
This also confirms that the active escape does not depend on the height of the barrier.

%%%%%%%%%%%%%%%%%%%%%%%%%%%%%%%%%%%%%%%%%%%%%%%%%%%%%%%%%%%%%%%%%%%%%%%%%%%
\section{Two-particle active escape-problem\label{sec:twoparticle}}
%%%%%%%%%%%%%%%%%%%%%%%%%%%%%%%%%%%%%%%%%%%%%%%%%%%%%%%%%%%%%%%%%%%%%%%%%%%
In this section, we consider the dynamics~\eqref{eq:activedynamics} with $N=2$ particles to understand the impact of the inter-particle interactions on the escape process in the double-well potential~\eqref{eq:doublewell}.
For this reason, we kept constant the parameters of the potential and the swim velocity, $v_0$, whose roles have been already analyzed and understood in the previous section.

As in the single-particle case, we first describe the regime of small $\tau$, 
where we expect the system to behave as a passive one with an effective potential, and  then the large $\tau$ regime.

%==========================================================================
\subsection{Small persistence regime}
%==========================================================================

%---------------------------- Fig.3 -------------------------------------
\begin{figure}
\includegraphics[width=0.95\columnwidth]{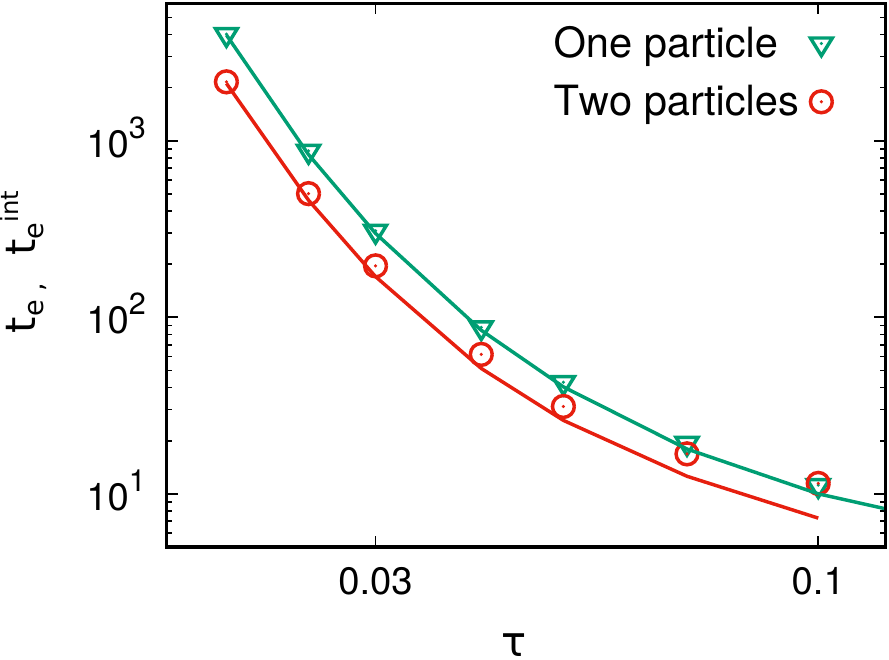}
\caption{Average escape time, $t_e$ (one particle) and $t_e^int$ for the one-particle system (green triangles) and the two-particles system (red circles), as a function of the persistence time $\tau$.  Colored solid lines are obtained by the theoretical predictions~\eqref{eq:kramersUCNA} (green line) and by Eq.~\eqref{eq:kramersUCNA} with $H\to H_{eff}$, where $H_{eff}$ is obtained by Eq.~\eqref{eq:Heff}.
The other parameters of the simulations are: $v_0=5$, $W_0=1$, $a=2$, $\epsilon=1$, $\sigma=1$.
\label{fig:smalltau_mod}}
\end{figure}
%------------------------------------------------------------------------

%---------------------------- Fig.4 -------------------------------------
\begin{figure*}[t]
\includegraphics[width=0.95\textwidth]{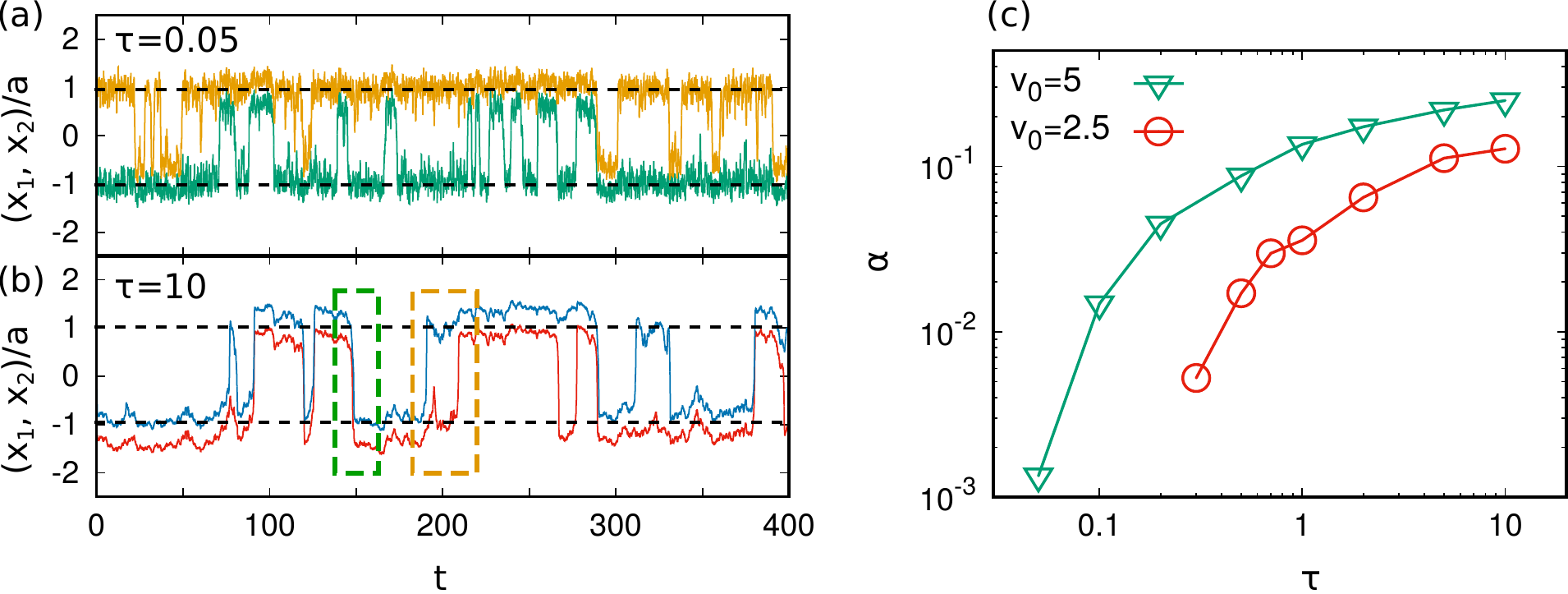}
\caption{Panels (a) and (b): time trajectories of the two particles (normalized by $a$), namely $x_1(t)/a$ (green in panel (a) and red in panel (b)) and $x_2(t)/a$ (yellow in panel (a) and blue in panel (b)). Panel (a) and (b) are obtained with $\tau=0.05, 10$, respectively, with $v_0=5$. In both panels the two dashed black lines mark the positions of the minima. The green dashed square is drawn in correspondence of a simultaneous jump while the yellow one in correspondence of two independent jumps.
Panel (c): fraction of simultaneous jump, $\alpha$, as a function of $\tau$ for two different values of the swim velocity, $v_0=2.5, 5$.
The other parameters of the simulations are: $W_0=1$, $a=2$, $\epsilon=1$, $\sigma=1$.
\label{fig:trajdoublejumps}}
\end{figure*}
%----------------------------------------------------------------------------
In analogy with Eq.~\eqref{eq:HamiltonianUCNA}, we can express   
the joint probability distribution $p_2(x_1, x_2)$ for the position of the two particles in the small-$\tau$ regime, as 
\begin{equation}
\label{eq:two-particles_distribution}
p_2(x_1, x_2) \propto \exp[-H_{tot}(x_1, x_2)/D_a\gamma] \,,
\end{equation} 
where $H_{tot}(x_1, x_2)$ is an effective Hamiltonian which can be decomposed as:
\begin{equation}
\label{eq:total_H}
H_{tot}(x_1, x_2)= H(x_1)+ H(x_2) + H_{int}(x_1, x_2) \,.
\end{equation}
The term $H$ is the single-particle effective Hamiltonian (which depends only on the external potential and its derivatives) already introduced in Eq.~\eqref{eq:HamiltonianUCNA}, while $H_{int}$ is the interaction Hamiltonian (containing the inter-particle potential, $U$) that reads:
\begin{equation}
\label{eq:UCNA_two}
\begin{aligned}
H_{int}&\approx \frac{\tau}{\gamma} \biggl[\left(\partial_{x_1} U \right)^2 + \left(\partial_{x_2} U \right)^2 + 2(\partial_{x_1} U) \partial_{x_1} W(x_1)\\
&+ 2(\partial_{x_2} U) \partial_{x_2} W(x_2)  \biggr]
 - D_a \gamma\log{\left( 1+2 \frac{\tau}{\gamma} \partial^2_{x_1} U \right)}\,.
\end{aligned}
\end{equation}
Notice that the last term follows from the approximation of the determinant of the Hessian matrix appearing in the UCNA distribution. For further details about this  result in the interacting (two-particles) case, see Ref.~\cite{marconi2016effective}.
We remark that $H_{int}$ not only depends on the interaction potential (as usual in passive systems) but also on the external potential and their derivatives which produce the effective attraction qualitatively responsible for cluster formation and motility induced phase separation~\cite{farage2015effective} (see also Ref.~\cite{rein2016applicability} for the parameter range for the application of such a method).

In virtue of these analytical results for $p_2$, it is possible to find an effective description for a tagged particle (namely, particle 1) by integrating out the coordinate of the second particle, $x_2$, in Eq.~\eqref{eq:two-particles_distribution}. 
% with the total Hamiltonian~\eqref{eq:total_H}.
Applying this procedure, we obtain the expression for the single-particle marginal probability distribution, $p_1(x_1)$, from which the single-particle effective Hamiltonian is derived by taking the logarithm, as follows:
\begin{equation}
\begin{aligned}
\label{eq:Heff}
&H_{\mathrm{eff}}(x_1)=H(x_1) \\
&- D_a \gamma\log{\int dx_2 \exp{\left(-\frac{H_{int}(x_1, x_2)+H(x_2)}{D_a\gamma}\right)} } \,,
\end{aligned}
\end{equation}
where we recall that $D_a=v_0^2 \tau$.
At this stage, Kramers' formula~\eqref{eq:kramersUCNA} can be easily applied by replacing $H$ with $H_{\mathrm{eff}}$ to derive an analytical expression for the effective escape time  $t_e^{\mathrm{int}}$, for the tagged particle in the interacting system.   

In Fig.~\ref{fig:smalltau_mod}, $t_e^{\mathrm{int}}$ is numerically studied as a function of $\tau$ (only small values of $\tau$ are shown) and the results are compared with the Kramers-like theoretical prediction.
As in the one-particle case, the agreement is fairly good for the smaller values of $\tau$, while the theoretical prediction underestimates the numerical value of $t_e^{\mathrm{int}}$ when $\tau$ is increased.
Fig.~\ref{fig:smalltau_mod} also reports the comparison between $t_e^{\mathrm{int}}$ and $t_e$ (the escape time from the double-well potential in the non-interacting case). 
As it occurs in passive systems, we see that $t_e^{\mathrm{int}} < t_e$. This result can be easily explained because the interaction potential decreases the effective potential barrier of the single-particle. Indeed, when the particles are placed in different wells, the escape follows the rules of the non-interacting problem. Instead, when the particles are placed in the same well, for instance the left one, the right particle can escape more easily with respect to a non-interacting particle, because it is roughly placed at $x_m^{int}\approx -a +\sigma$ (see also Fig.~\ref{fig:trajdoublejumps}~(a)) and, thus, the single-particle effective potential barrier is reduced with respect to the bare value $W_b = W_0 a^4/4$. 
We also observe that, as the persistence is increased, the difference between $t_e$ and $t_e^{\mathrm{int}}$ reduces until $t_e^{\mathrm{int}}\approx t_e$.
This occurs because the interaction Hamiltonian, $H_{int}$, contains effective attractions terms whose relevance increases when the persistence time grows~\cite{marconi2016effective}.
These effective attractive terms hinder the escape from a well, similarly to the real attraction in passive systems, see e.g. Ref.~\cite{asfaw2012exploring}, for the case of two particles interacting via a harmonic force, forming a dimer. 

%==================================================================
\subsection{Large persistence regime}
%==================================================================
As we expect, the UCNA prediction~\eqref{eq:UCNA_two} cannot work in the large persistence regime like in the one-particle case. In the absence of a theoretical picture, we resort to numerically study the active escape properties.
Before delving into the study of the mean escape time, we focus on the phenomenology of the escape process to understand the difference between small and large persistence regimes.
Fig.~\ref{fig:trajdoublejumps}~(a) and~(b) show the single-particle trajectories, namely the positions of the two particles, $x_1(t)$ and $x_2(t)$ normalized by $a$, as a function of time. 
In the two panels, two different values of $\tau$ are considered as illustrative cases for the two regimes.
In the small persistence regime (Fig.~\ref{fig:trajdoublejumps}~(a)), it is not surprising that the two particles perform uncorrelated jumps so that each particle independently escapes from a well as it occurs in a system of two passive particles.
Instead, in the large persistence regime, the escape process is quite different, as seen in Fig.~\ref{fig:trajdoublejumps}~(b), because, among the escape events, there is a non-negligible fraction jump involving both particles, whereby they cross the barrier almost simultaneously (see dashed green rectangles).
We refer to these events as correlated jumps.
Let us suppose that the particles are both in the left well, $x_1<x_2<0$.  
The active particle ``$1$'', which is farther from the barrier $x=0$, could be able to drag the particle ``$2$'' towards $x=0$, forcing its escape even if the active force of ``$2$'' is smaller than $f_{M}$, thus producing a simultaneous jump.
To make this picture more quantitative, in Fig.\ref{fig:trajdoublejumps}~(c), we measure the fraction, $\alpha$, of correlated jumps occurring in a long-time simulation run as a function of $\tau$ for two different values of $v_0$. 
This fraction is defined as $\alpha=n_s/n_{tot}$ being $n_s$ the number of simultaneous jumps and $n_{tot}$ the total number of jumps occurring in the time-window of the run.
As expected, $\alpha$ is an increasing (monotonic) function of $\tau$: the larger is the persistence, the more probable is the occurrence of a correlated jump.
Interestingly, depending on the value of $v_0$, the fraction $\alpha$ could reach also large values, so that even the 10\% or 20\% of the escape events could be simultaneous. 
This correlated-escape scenario has not a passive counterpart, since in that case the probability of observing a simultaneous jump is always negligible. However, correlated jumps have been observed in granular systems, where dissipative collisions determine a kind of ``effective'' attraction similar to that discussed in this paper~\cite{cecconi2003noise}.

%---------------------------- Fig.5 -------------------------------------
\begin{figure}
\includegraphics[width=0.95\columnwidth]{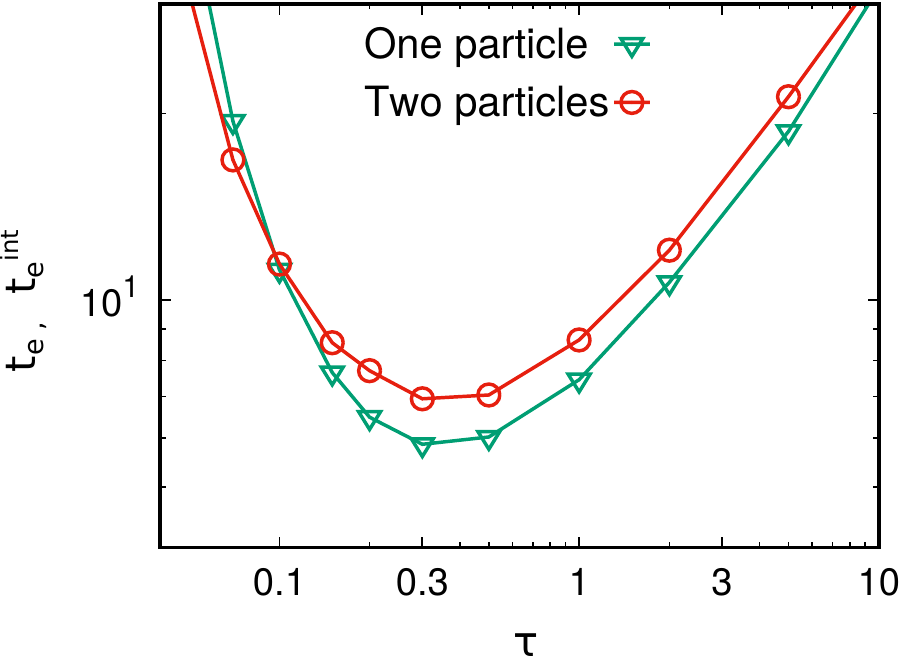}
\caption{Average escape time, $t_e$ (one particle) and $t_e^int$ for the one-particle system (green triangles) and the two-particles system (red circles), as a function of the persistence time $\tau$. Solid lines are guides for the eyes.
The other parameters of the simulations are: $v_0=5$, $W_0=1$, $a=2$, $\epsilon=1$, $\sigma=1$.
\label{fig:smalltau_two_large_mod}}
\end{figure}
%------------------------------------------------------------------------

Finally, Fig.~\ref{fig:smalltau_two_large_mod} shows the average escape times, $t_e^{\mathrm{int}}$ and $t_e$, for the interacting and non-interacting cases, respectively, as a function of $\tau$, exploring also values of $\tau$ outside the applicability of UCNA.
Both the escape times decreases until a minimum is reached and, then, for further values of $\tau$, they monotonically increase.
We remark that, at variance with the small-$\tau$ regime, now we observe $t_e^{\mathrm{int}}>t_e$. 
This scenario could be explained by the well-known slow-down due to the interplay between the active and repulsive forces, which concur to produce an effect qualitatively similar to an effective attraction. This mechanism hinders the escape process with respect to the non-interacting case.

%%%%%%%%%%%%%%%%%%%%%%%%%%%%%%%%%%%%%%%%%%%%%%%%%%%%%%%%%%%%%%%%%%%%%%%%%%%%%%%%%%
\section{Conclusion}
%%%%%%%%%%%%%%%%%%%%%%%%%%%%%%%%%%%%%%%%%%%%%%%%%%%%%%%%%%%%%%%%%%%%%%%%%%%%%%%%%%
In this paper, we have studied the escape properties of active particles - using the active Ornstein-Uhlenbeck model - confined in a double-well potential, with a particular focus on the mean escape time from a well.
At first, we have investigated the escape of a single-particle considering a wide range of persistence regimes of the active force. 
Interestingly, we have observed the existence of an optimum value of the persistence time (at fixed external potential), which minimizes the mean escape time. We have also developed a theoretical explanation for the behavior at small-persistence regime, combining the Unified Colored Noise approximation (UCNA) with a Kramers-like theory. 
We have also proposed a simple theoretical argument to explain the linear growth of the mean escape time in the large persistence regime.

As a second step, we have considered a system of two active particles interacting via a repulsive potential.
Also this case, we derive a theoretical prediction for the average escape time holding in the small persistence regime: as expected for passive particles, the escape is favored in the interacting system because each particle behaves as if was affected by an effective barrier lower than the barrier of the external potential.
Interestingly, in the large persistence regime, we observe the opposite: the interplay between the active force and the repulsive interaction induces an effective attraction between the particles which hinder the escape process in the two-particle systems.
In addition, we outline the peculiar properties of the two-particle active escape: in the regime of large persistence, we observe a large fraction of simultaneous jumps (which occurs when the two particles jump together) that does not have a passive counterpart.

Recently, Br\"{u}ckner et. al. have performed intriguing experiments of strongly confined active particles: they consider epithelial (even cancerous) cells in a simple geometry consisting of two adhesive sites connected by a thin constriction~\cite{bruckner2019stochastic, bruckner2020disentangling, fink2020area}. 
The cell, which has a certain degree of persistence, migrates from a site to the other and seems to behave as if it is subject to a double-well potential along with the travel direction.
The results of these experiments, in particular the position-velocity phase space and the features of the jump mechanism, are in qualitative agreement with those obtained by numerical simulations using the active Ornstein-Uhlenbeck model  in a double-well potential which shows a bifurcation-like scenario in the proximity of the potential maximum~\cite{caprini2019activeescape}.
By such an analogy, we believe that our work could stimulate future experimental and numerical studies focused on the escape properties (i.e. the time that a cell needs to cross a thin constriction) and shed light on the jump properties of such an experimental system.

\subsection*{Acknowledgements}
LC, FC and UMBM acknowledge support from the MIUR PRIN 2017 project 201798CZLJ. LC acknowledges support from the Alexander Von Humboldt foundation.

\subsection*{Data availability}
The data that support the findings of this study are available from the corresponding author upon reasonable request.

\appendix
%%%%%%%%%%%%%%%%%%%%%%%%%%%%%%%%%%%%%%%%%%%%%%%%%%%%%%%%%%%%%%%%%%%%
\section{Derivation of Eq.~\eqref{eq:prediction_te_largetau}\label{app:derivation}}
%%%%%%%%%%%%%%%%%%%%%%%%%%%%%%%%%%%%%%%%%%%%%%%%%%%%%%%%%%%%%%%%%%%%
In this Appendix, we derive the expression~\eqref{eq:prediction_te_largetau} for the escape time $t_e$, holding in the limit of $\tau\to \infty$. As also discussed in Sec.~\ref{sec:oneparticle}, in this regime of $\tau$, $t_e$ can be determined by the mean first passage time taken by the active force $f^a$ to overcome the threshold $|f_{M}|$ (namely, the maximal force experienced by a particle in climbing the barrier).
%Eq.~\eqref{eq:prediction_te_largetau} is the mean first-arrival time of the Ornstein-Uhlenbeck (OU) process starting from the generic value $f$ and reaching the boundary $f_{\mathrm{max}}$ which is considered absorbing. 
According to the first passage theory (see Ref.~\cite{Redner2001guide}), this time is  related to the survival probability, $S(f^a,t)$, by the integral~\cite{gardiner1985handbook}:
$$
T(f^a) = \int_{0}^{\infty}\!\!dt\;S(f^a,t)\,. 
$$
By definition, $S(f^a,t)$ is the probability that the process has not yet reached $f_{\mathrm{max}}$ at time $t$. $S(f^a,t)$ is known to satisfy the backward Fokker-Planck equation~\cite{gardiner1985handbook}, that for the OU process reads
\begin{equation}
\dfrac{\partial S}{\partial t} = 
-\dfrac{f^a}{\tau}\dfrac{\partial S}{\partial f^a} +
\dfrac{(\gamma v_0)^2}{\tau}\dfrac{\partial^2 S}{\partial (f^a)^2},
\label{eq:Survive}
\end{equation}
with the boundary conditions $S(f_{\mathrm{max}},t)=0$. 
In the following, for the sake of concision, we set $a=(\gamma v_0)^2$.
To obtain a differential equation for $T(f^a)$, it is sufficient to integrate Eq.~\eqref{eq:Survive} in the interval $0\le t < \infty$, and taking into account that $S(f^a,\infty) = 0$ and $S(f^a,0) = 1$, we get:
\begin{equation}
-f^a\dfrac{\partial T}{\partial f^a} +
 a \dfrac{\partial^2 T}{\partial x^2} =-\tau \,,
\label{eq:MFPT}
\end{equation}
that has to be solved with the boundary conditions $T(f_{\mathrm{max}}) = 0$ and $T'(-\infty) = 0$.
The first condition states that a process started at the boundary $f_{M}$ is instantaneously absorbed and the second one that $f^a=-\infty$ acts as a reflecting barrier, since very large $f^a$-values are practically inaccessible due to the quadratic form of the potential, $f^2/2\tau$. 
The solution of Eq.~\eqref{eq:MFPT} can be obtained by quadrature, setting $T'(f) = w(f)$, and reads:
$$
T(f^a) = \dfrac{\tau}{a} \int_{f^a}^{f_{M}}\!\!\!
dx e^{x^2/(2 a)} 
\int_{-\infty}^{x}\!\!\! dy e^{-y^2/(2a)} \,.
$$
After performing the integration on $y$, one obtains:
\begin{equation}
\label{eq:app_Tfint}
T(f^a) = \tau\;\sqrt{\dfrac{\pi}{2a}} \int_{f^a}^{f_{M}}\!\!\!
dx\;e^{x^2/(2a)} \bigg(1+\mbox{Erf}\bigg[\dfrac{x}{\sqrt{2a}}\bigg]\bigg) \,.
\end{equation}
Finally, by evaluating the expression~\eqref{eq:app_Tfint} at $f^a=0$ and after a change of variable in the integration, we obtain Eq.~\eqref{eq:prediction_te_largetau}, since $T(f^a=0)=t_e$.

%\end{widetext}

%\bibliographystyle{mdpi}
%\bibliographystyle{rsc} %the RSC's .bst file
\bibliographystyle{rsc} %the RSC's .bst file

\bibliography{active.bib}

%%%%%%%%%%%%%%%%%%%%%%%%%%%%%%%%%%%%%%%%%%
%% optional
%\sampleavailability{Samples of the compounds ...... are available from the authors.}

\end{document}